\documentclass{ws-ijmpa}

\begin{document}

\def\bz{{B^0}}
\def\bzb{{\overline{B}{}^0}}
\def\bp{{B^+}}
\def\bm{{B^-}}
\def\kl{K_L^0}
\def\dE{{\Delta E}}
\def\mb{{M_{\rm bc}}}
\def\Dt{\Delta t}
\def\Dz{\Delta z}
\def\fol{f_{\rm ol}}
\def\fsig{f_{\rm sig}}
\newcommand{\sinbb}{{\sin2\phi_1}}
 
\newcommand{\ra}{\rightarrow}
\newcommand{\myindent}{\hspace*{2cm}}  
\newcommand{\fCP}{f_{CP}}
\def\fcp{\fCP}
\newcommand{\ftag}{f_{\rm tag}}
\newcommand{\zCP}{z_{CP}}
\newcommand{\tCP}{t_{CP}}
\newcommand{\ttag}{t_{\rm tag}}
\newcommand{\cala}{{\cal A}}
\newcommand{\cals}{{\cal S}}
\newcommand{\dm}{\Delta m_d}
\newcommand{\dmd}{\dm}
\def\taubz{{\tau_\bz}}
\def\taubp{{\tau_\bp}}
\def\ks{{K_S^0}}
\newcommand{\btosqq}{b \to s\overline{q}q}
\newcommand{\btosss}{b \to s\overline{s}s}
\newcommand*{\dwl}{\ensuremath{{\Delta w_l}}}
\newcommand*{\fq}{\ensuremath{q}}
\def\kz{{K^0}}
\def\kp{{K^+}}
\def\km{{K^-}}
\def\fzero{{f_0(980)}}
\def\pip{{\pi^+}}
\def\pim{{\pi^-}}
\def\piz{{\pi^0}}
\def\kstarz{{K^{*0}}}
\def\kstarp{{K^{*+}}}
\def\kstarm{{K^{*-}}}
\def\kstarpm{{K^{*\pm}}}
\def\kl{{K_L^0}}
\def\bbar{{\overline{B}}}
\def\ufs{{\Upsilon(4S)}}
\def\nev{{N_{\rm ev}}}
\def\nsig{{N_{\rm sig}}}
\def\Nev{\nev}
\def\nsigmc{{N_{\rm sig}^{\rm MC}}}
\def\nbkg{{N_{\rm bkg}}}

\def\jpsi{{J/\psi}}
\def\dminus{{D^-}}
\def\dplus{{D^+}}
\def\dsm{{D^{*-}}}
\def\dpm{{D^{*+}}}
\def\rhop{{\rho^+}}
\def\rhom{{\rho^-}}
\def\rhoz{{\rho^0}}
\def\dzero{{D^0}}
\def\dzerob{{\overline{D}{}^0}}

\def\dzb{{\overline{D}{}^0}}
\newcommand{\dslnu}{D^{*-}\ell^+\nu}
\newcommand{\bzdslnu}{\bz \to \dslnu}
\newcommand*{\fflv}{\ensuremath{{f_\textrm{flv}}}}
\newcommand{\thetabdl}{\theta_{B,D^*\ell}}
\newcommand{\cosbdl}{\cos\thetabdl}

\def\sperp{{S_{\perp}}}
\def\lsig{{\cal L}_{\rm sig}}
\def\lbkg{{\cal L}_{\rm bkg}}
\def\rsigbkg{{\cal R}_{\rm s/b}}
\def\calf{{\cal F}}
\def\rkpi{{\cal R}_{K/\pi}}

\def\mgg{M_{\gamma\gamma}}
\def\ppizcms{p_\piz^{\rm cms}}

\def\pbstar{p_B^{\rm cms}}

\newcommand*{\eeff}{\ensuremath{\epsilon_\textrm{eff}}}
\def\egcms{{E_\gamma^{\rm cms}}}

\def\lnsig{{\cal L}_{N_{\rm sig}}}
\def\lzero{{\cal L}_0}

\def\acpraw{{A_{CP}^{\rm raw}}}

\def\efftot{{0.30\pm 0.01}}

\def\sinbbWA{+0.726}
\def\sinbbERR{0.037}
\def\sinbbWAResult{\sinbbWA\pm\sinbbERR}

\def\SphiksResPrv{-0.96\pm0.50^{+0.09}_{-0.11}}
\def\AphiksResPrv{-0.15\pm0.29\pm0.07}
\def\SetapksResPrv{+0.43\pm0.27\pm0.05}
\def\AetapksResPrv{-0.01\pm0.16\pm0.04}
\def\SkpkmksResPrv{-0.51\pm0.26\pm0.05}
\def\AkpkmksResPrv{-0.17\pm0.16\pm0.04}

\def\Nevjpsikspm{xxxx} 
\def\Pjpsikspm{0.xx} \def\Nsigjpsikspm{xxxx\pm xx}
 \def\Nevjpsikszz{xxx} 
\def\Pjpsikszz{0.xx} \def\Nsigjpsikszz{xxx\pm xx} 
  \def\Nevjpsikl{xxxx}     
  \def\Pjpsikl{0.xx}    \def\Nsigjpsikl{xxxx\pm xx}
    \def\Nevphiks{221}       
   \def\Pphiks{0.63}     \def\Nsigphiks{139 \pm 14}
    \def\Nevphikl{207}       
   \def\Pphikl{0.17}     \def\Nsigphikl{36 \pm 15}
\def\Nevkpkmks{718}     
  \def\Pkpkmks{0.56}    \def\Nsigkpkmks{399\pm 28}
\def\NevkspizH{298}      
   \def\PkspizH{0.55}     \def\NsigkspizH{168\pm 16}
\def\NevkspizL{499}      
   \def\PkspizL{0.17}     \def\NsigkspizL{83\pm 18}
\def\Nevkstarzgm{92} 
\def\Pkstarzgm{0.65}  \def\Nsigkstarzgm{57 \pm 9}
\def\Nevetapks{842}   
  \def\Petapks{0.61}    \def\Nsigetapks{512 \pm 27}
\def\Nevomegaks{56}  
 \def\Pomegaks{0.56}    \def\Nsigomegaks{31 \pm 7}
\def\Nevfzeroetcks{178}  
 \def\Pfzeroetcks{0.58}    \def\Nsigfzeroetcks{102 \pm 12}
 \def\Pfzeroks{0.53}    \def\Nsigfzeroks{94 \pm 14}
%
\def\SjpsikzVal{+0.666}   \def\SjpsikzStat{0.046}   \def\SjpsikzSyst{x.xx}
\def\AjpsikzVal{+0.023}   \def\AjpsikzStat{0.031}   \def\AjpsikzSyst{x.xx}
\def\SphikzVal{+0.06}    \def\SphikzStat{0.33}    \def\SphikzSyst{0.09}
\def\AphikzVal{+0.08}    \def\AphikzStat{0.22}    \def\AphikzSyst{0.09}
\def\SphiksVal{-x.xx}    \def\SphiksStat{x.xx}    \def\SphiksSyst{x.xx}
\def\AphiksVal{-x.xx}    \def\AphiksStat{x.xx}    \def\AphiksSyst{x.xx}
\def\SphiklVal{-x.xx}    \def\SphiklStat{x.xx}    \def\SphiklSyst{x.xx}
\def\AphiklVal{-x.xx}    \def\AphiklStat{x.xx}    \def\AphiklSyst{x.xx}
\def\SkpkmksVal{-0.49}   \def\SkpkmksStat{0.18}   \def\SkpkmksSyst{0.04}
\def\AkpkmksVal{-0.08}   \def\AkpkmksStat{0.12}   \def\AkpkmksSyst{0.07}
\def\SkpkmksFCP{^{+0.18}_{-0.00}}
\def\SkspizVal{+0.30}    \def\SkspizStat{0.59}    \def\SkspizSyst{0.11}
\def\AkspizVal{-0.12}    \def\AkspizStat{0.20}    \def\AkspizSyst{0.07}
\def\SkstarzgmVal{-0.79} \def\SkstarzgmStat{^{+0.63}_{-0.50}} \def\SkstarzgmSyst{0.10}
\def\AkstarzgmVal{-0.00} \def\AkstarzgmStat{0.38} \def\AkstarzgmSyst{x.xx}
\def\SetapksVal{+0.65}   \def\SetapksStat{0.18}   \def\SetapksSyst{0.04}
\def\AetapksVal{-0.19}   \def\AetapksStat{0.11}   \def\AetapksSyst{0.05}
\def\SomegaksVal{+0.75}  \def\SomegaksStat{0.64}  \def\SomegaksSyst{^{+0.13}_{-0.16}}
\def\AomegaksVal{+0.26}  \def\AomegaksStat{0.48}  \def\AomegaksSyst{0.15}
\def\SfzeroksVal{+0.47}  \def\SfzeroksStat{0.41}  \def\SfzeroksSyst{0.08}
\def\AfzeroksVal{-0.39}  \def\AfzeroksStat{0.27}  \def\AfzeroksSyst{0.08}
\def\SbsqqVal{+0.43} \def\SbsqqErr{^{+0.12}_{-0.11}}

\def\SjpsikzResult{\SjpsikzVal\pm\SjpsikzStat\pm\SjpsikzSyst}
\def\SjpsikzResultSS
  {\SjpsikzVal\pm\SjpsikzStat\mbox{(stat)}\pm\SjpsikzSyst\mbox{(syst)}}
\def\AjpsikzResult{\AjpsikzVal\pm\AjpsikzStat\pm\AjpsikzSyst}
\def\AjpsikzResultSS
  {\AjpsikzVal\pm\AjpsikzStat\mbox{(stat)}\pm\AjpsikzSyst\mbox{(syst)}}
\def\SphikzResult{\SphikzVal\pm\SphikzStat\pm\SphikzSyst}
\def\SphikzResultSS
  {\SphikzVal\pm\SphikzStat\mbox{(stat)}\pm\SphikzSyst\mbox{(syst)}}
\def\AphikzResult{\AphikzVal\pm\AphikzStat\pm\AphikzSyst}
\def\AphikzResultSS
  {\AphikzVal\pm\AphikzStat\mbox{(stat)}\pm\AphikzSyst\mbox{(syst)}}
\def\SphiksResult{\SphiksVal\pm\SphiksStat\pm\SphiksSyst}
\def\SphiksResultSS
  {\SphiksVal\pm\SphiksStat\mbox{(stat)}\pm\SphiksSyst\mbox{(syst)}}
\def\AphiksResult{\AphiksVal\pm\AphiksStat\pm\AphiksSyst}
\def\AphiksResultSS
  {\AphiksVal\pm\AphiksStat\mbox{(stat)}\pm\AphiksSyst\mbox{(syst)}}
\def\SphiklResult{\SphiklVal\pm\SphiklStat\pm\SphiklSyst}
\def\SphiklResultSS
  {\SphiklVal\pm\SphiklStat\mbox{(stat)}\pm\SphiklSyst\mbox{(syst)}}
\def\AphiklResult{\AphiklVal\pm\AphiklStat\pm\AphiklSyst}
\def\AphiklResultSS
  {\AphiklVal\pm\AphiklStat\mbox{(stat)}\pm\AphiklSyst\mbox{(syst)}}
\def\SkpkmksResult{\SkpkmksVal\pm\SkpkmksStat\pm\SkpkmksSyst}
\def\SkpkmksResultSS
  {\SkpkmksVal\pm\SkpkmksStat\mbox{(stat)}\pm\SkpkmksSyst\mbox{(syst)}}
\def\AkpkmksResult{\AkpkmksVal\pm\AkpkmksStat\pm\AkpkmksSyst}
\def\AkpkmksResultSS
  {\AkpkmksVal\pm\AkpkmksStat\mbox{(stat)}\pm\AkpkmksSyst\mbox{(syst)}}
\def\SkspizResult{\SkspizVal\pm\SkspizStat\pm\SkspizSyst}
\def\SkspizResultSS
  {\SkspizVal\pm\SkspizStat\mbox{(stat)}\pm\SkspizSyst\mbox{(syst)}}
\def\AkspizResult{\AkspizVal\pm\AkspizStat\pm\AkspizSyst}
\def\AkspizResultSS
  {\AkspizVal\pm\AkspizStat\mbox{(stat)}\pm\AkspizSyst\mbox{(syst)}}
\def\SkstarzgmResult{\SkstarzgmVal\SkstarzgmStat\pm\SkstarzgmSyst}
\def\SkstarzgmResultSS
  {\SkstarzgmVal\SkstarzgmStat\mbox{(stat)}\pm\SkstarzgmSyst\mbox{(syst)}}
\def\AkstarzgmResult{\AkstarzgmVal\pm\AkstarzgmStat}
\def\AkstarzgmResultSS
  {\AkstarzgmVal\pm\AkstarzgmStat\mbox{(stat)}\pm\AkstarzgmSyst\mbox{(syst)}}
\def\SetapksResult{\SetapksVal\pm\SetapksStat\pm\SetapksSyst}
\def\SetapksResultSS
  {\SetapksVal\pm\SetapksStat\mbox{(stat)}\pm\SetapksSyst\mbox{(syst)}}
\def\AetapksResult{\AetapksVal\pm\AetapksStat\pm\AetapksSyst}
\def\AetapksResultSS
  {\AetapksVal\pm\AetapksStat\mbox{(stat)}\pm\AetapksSyst\mbox{(syst)}}
\def\SomegaksResult{\SomegaksVal\pm\SomegaksStat\pm\SomegaksSyst}
\def\SomegaksResultSS
  {\SomegaksVal\pm\SomegaksStat\mbox{(stat)}\pm\SomegaksSyst\mbox{(syst)}}
\def\AomegaksResult{\AomegaksVal\pm\AomegaksStat\pm\AomegaksSyst}
\def\AomegaksResultSS
  {\AomegaksVal\pm\AomegaksStat\mbox{(stat)}\pm\AomegaksSyst\mbox{(syst)}}
\def\SfzeroksResult{\SfzeroksVal\pm\SfzeroksStat\pm\SfzeroksSyst}
\def\SfzeroksResultSS
  {\SfzeroksVal\pm\SfzeroksStat\mbox{(stat)}\pm\SfzeroksSyst\mbox{(syst)}}
\def\AfzeroksResult{\AfzeroksVal\pm\AfzeroksStat\pm\AfzeroksSyst}
\def\AfzeroksResultSS
  {\AfzeroksVal\pm\AfzeroksStat\mbox{(stat)}\pm\AfzeroksSyst\mbox{(syst)}}
\def\SbsqqResult{\SbsqqVal\SbsqqErr}

\markboth{Alexei Garmash}
{CP Violation in $b\to s$ Decays and New Physics Phases}

%
\catchline{}{}{}{}{}
%

\title{CP Violation in $b\to s$ Decays and New Physics Phases }

\author{\footnotesize ALEXEI GARMASH\\
(for the Belle Collaboration)}

\address{Department of Physics, Princeton University \\
Jadwin Hall, Princeton New Jersey 08544, U.S.A.}

\maketitle


\begin{abstract}
  We present new measurements of $CP$-violation parameters
  in 
  $\bz\to$
  $\phi\kz$, 
  $\kp\km\ks$,
  $\fzero\ks$,
  $\eta'\ks$, 
  $\omega\ks$, 
  $\ks\piz$, 
  and
  $\kstarz\gamma~(\kstarz\to\ks\piz)$
  decays
  based on a sample of $275\times 10^6$ $B\bbar$ pairs
  collected at the $\ufs$ resonance with
  the Belle detector at the KEKB energy-asymmetric $e^+e^-$ collider.
  One neutral $B$ meson is fully reconstructed in
  one of the specified decay channels,
  and the flavor of the accompanying $B$ meson is identified from
  its decay products.
  $CP$-violation parameters for each of the decay 
  modes are obtained from the asymmetries in the distributions of
  the proper-time intervals between the two $B$ decays.
  All results are preliminary.

\end{abstract}

\section{Introduction}	

$CP$ violation in the flavor-changing $b \to s$ transitions
is sensitive to physics at a very high-energy scale.
Theoretical studies indicate that
large deviations from standard model (SM) expectations
are allowed for time-dependent $CP$ asymmetries in
$\bz$ meson decays~\cite{bib:lucy}.
Belle's previous measurement of the $\bz\to \phi\ks$ decay,
which is dominated by the $\btosss$ transition,
yielded a value that differs from the SM expectation 
by 3.5 standard deviations~\cite{Abe:2003yt}.
Measurements with a larger data sample are required to 
elucidate this difference. It is also essential to examine
additional modes that
may be sensitive to the same $b \to s$ penguin amplitude.

In the SM, $CP$ violation arises from an irreducible phase,
the Kobayashi-Maskawa (KM) phase, in the weak-interaction
quark-mixing matrix. In particular, the SM predicts $CP$
asymmetries in the time-dependent rates for $\bz$ and
$\bzb$ decays to a $CP$ eigenstate $\fCP$.
In the decay chain $\Upsilon(4S)\to \bz\bzb \to f_{CP}f_{\rm tag}$,
where one of the $B$ mesons decays at time $t_{CP}$ to a 
final state $f_{CP}$ 
and the other decays at time $t_{\rm tag}$ to a final state  
$f_{\rm tag}$ that distinguishes between $B^0$ and $\bzb$, 
the decay rate has a time dependence
given by
\begin{equation}
\label{eq:psig}
{\cal P}(\Delta{t}) = 
\frac{e^{-|\Delta{t}|/{\taubz}}}{4{\taubz}}
\biggl\{1 + \fq\cdot 
\Bigl[ \cals\sin(\dmd\Delta{t})
   + \cala\cos(\dmd\Delta{t})
\Bigr]
\biggr\}.
\end{equation}
Here $\cals$ and $\cala$ are $CP$-violation parameters,
$\taubz$ is the $B^0$ lifetime, $\dmd$ is the mass difference 
between the two $B^0$ mass
eigenstates, $\Delta{t}$ = $t_{CP}$ $-$ $t_{\rm tag}$, and
the $b$-flavor charge $\fq$ = +1 ($-1$) when the tagging $B$ meson
is a $B^0$ ($\bzb$). To a good approximation,
the SM predicts $\cals = -\xi_f\sin 2\phi_1$, where $\xi_f = +1 (-1)$ 
corresponds to  $CP$-even (-odd) final states, and $\cala =0$
for both $b \to c\overline{c}s$ and $b \to s\overline{s}s$ transitions.
Possible contribution from $b\to u$ tree transition is estimated
to be in the range from few per cent to about ten per cent depending
on the final state~\cite{bib:b2u}.
In our analysis we neglect $b\to u$ contribution.

Another class of decays that proceed via the penguin
transition is radiative $b\to s\gamma$ decays.
Within the SM, the photon emitted from a $\bz$ ($\bzb$)
meson is dominantly right-handed (left-handed).
Therefore the polarization of the photon carries
information on the original $b$-flavor; the decay
is thus almost flavor-specific. The SM predicts
a small asymmetry $\cals \sim -2(m_s/m_b)\sinbb$, where $m_b$ ($m_s$) is
the $b$-quark ($s$-quark) mass~\cite{Atwood:1997zr}.
Any significant deviation from this expectation would be a
manifestation of physics beyond the SM.

Recently Belle and BaBar measured time-dependent $CP$ asymmetries in
$\bz \to J/\psi \ks$ and related decay modes, which are governed by
the $b \to c\overline{c}s$ transition, and
already determined $\sinbb$ rather precisely;
the present world average value is 
$\sinbb = \sinbbWAResult$~\cite{bib:HFAG}.
This serves as a firm reference point for the SM.

\section{Data Sample and Analysis Technique}

Our previous measurements
for $\bz \to \phi\ks$, $K^+K^-\ks$ and $\eta'\ks$
were based on a 140~fb$^{-1}$ data sample (DS-I)
with $152\times 10^6$ $B\bbar$ pairs~\cite{Abe:2003yt}.
In this report, we describe improved measurements 
incorporating an additional 113~fb$^{-1}$
data sample that contains $123\times 10^6$ $B\bbar$
pairs (DS-II) for a total of $275\times 10^6$ $B\bbar$ pairs.
Two inner detector configurations were used. A 2.0 cm radius beampipe
and a 3-layer silicon vertex detector (SVD-I) were used for DS-I,
while a 1.5 cm radius beampipe, a 4-layer
silicon detector (SVD-II) and a small-cell inner drift chamber were used for 
DS-II~\cite{Ushiroda}.

In this update we include additional $\phi\ks$, $\ks\to\pi^0\pi^0$
and  $\eta'\ks$, $\eta'\to\eta\pi^+\pi^-$, $\eta\to\pi^+\pi^-\pi^0$
subdecay modes that were not used in the previous analysis.
We also perform new measurements of $CP$ asymmetries for
the following $CP$-eigenstate $\bz$ decay modes:
$\bz\to \phi\kl$ and $\fzero\ks$ for $\xi_f = +1$;
$\bz\to\omega\ks$ and $\ks\piz$ for $\xi_f = -1$.
The decays $\bz\to\phi\ks$ and $\phi\kl$ are combined
in this analysis by redefining $\cals$ as $-\xi_f\cals$ to take
the opposite $CP$ parities into account, 
and are collectively called ``$\bz\to\phi\kz$''.
The $CP$ asymmetries for the decay $\bz\to\omega\ks$
are measured for the first time.
Finally, we also measure time-dependent $CP$ violation in
the decay $\bz\to\kstarz\gamma~(\kstarz\to\ks\piz)$,
which is not a $CP$ eigenstate but is sensitive to
physics beyond the SM. Yet another final state
that is not a $CP$ eigenstate is $K^+K^-\ks$.
We find that the $K^+K^-\ks$ state is primarily $\xi_f=+1$;
a measurement of the $\xi_f=+1$ fraction with DS-I gives
$1.03 \pm 0.15\mbox{(stat)}\pm0.05\mbox{(syst)}$~\cite{Abe:2003yt}.
In the following determination of $\cals$ and $\cala$,
we fix $\xi_f=+1$ for this mode.

We determine $\cals$ and $\cala$ for each mode by performing an unbinned
maximum-likelihood fit to the observed $\Dt$ distribution.
The probability density function expected for the signal
distribution,
is given by Eq.~(\ref{eq:psig}) incorporating
the effect of incorrect flavor assignment. The distribution is
convolved with the
proper-time interval resolution function,
which takes into account the finite vertex resolution. 

For the decays $\bz\to\ks\piz$ and $\kstarz\gamma~(\kstarz\to\ks\piz)$,
the $B$ vertex reconstruction technique is validated using
$\bz\to\jpsi\ks$ events where the $B$ decay vertex is reconstructed
with $\ks$ and the IP constraint rather than charged tracks from $\jpsi$.
A CP fit to the $\bz\to\jpsi\ks$ sample gives
$\cals_{\jpsi\ks} = +0.68\pm 0.10$(stat) and
$\cala_{\jpsi\ks} = +0.02\pm 0.04$(stat), which
are in good agreement with the world average values.
Thus, we conclude that 
the vertex resolution for the $\bz\to\ks\piz$ and
$\bz\to\kstarz\gamma~(\kstarz\to\ks\piz)$ decays
is well understood.
The only free parameters in the final fit are $\cals$ and $\cala$.

Event selections and analysis technique are described in
details in Ref.~\cite{bib:b2s-new} and references therein.
Here we only discuss the results.

\section{Results of {\boldmath $CP$} Asymmetry Measurements}

\begin{table}[t]
\tbl{Results of the fits to the $\Dt$ distributions.
Errors are statistical and systematic. We combine
$\bz\to\phi\ks$ and $\bz\to\phi\kl$ decays
to obtain $\cals_{\phi\kz}$ and $\cala_{\phi\kz}$.}
{\begin{tabular}{@{}lrll@{}} \toprule
\multicolumn{1}{c}{Mode} &  
SM expectation for $\cals$ &
\multicolumn{1}{c}{$\cals$} & 
\multicolumn{1}{c}{$\cala$} \\
 \colrule
$\phi\kz$   & $+\sinbb$   & $\SphikzResult$    & $\AphikzResult$    \\
$K^+K^-\ks$ & $-\sinbb$   & $\SkpkmksResult$   & $\AkpkmksResult$   \\
$\fzero\ks$ & $-\sinbb$   & $\SfzeroksResult$  & $\AfzeroksResult$  \\
$\eta'\ks$  & $+\sinbb$   & $\SetapksResult$   & $\AetapksResult$   \\
$\omega\ks$ & $+\sinbb$   & $\SomegaksResult$  & $\AomegaksResult$  \\
$\ks\piz$   & $+\sinbb$   & $\SkspizResult$    & $\AkspizResult$    \\
$\kstarz\gamma~(\kstarz\to\ks\piz)$ 
            & $-2(m_s/m_b)\sinbb$ 
                          & $\SkstarzgmResult$ & ~~~~~~$0.0$ (fixed) \\
 \botrule
\end{tabular}}
\label{tab:results}
\end{table}

\begin{figure}
\includegraphics[width=0.33\textwidth]{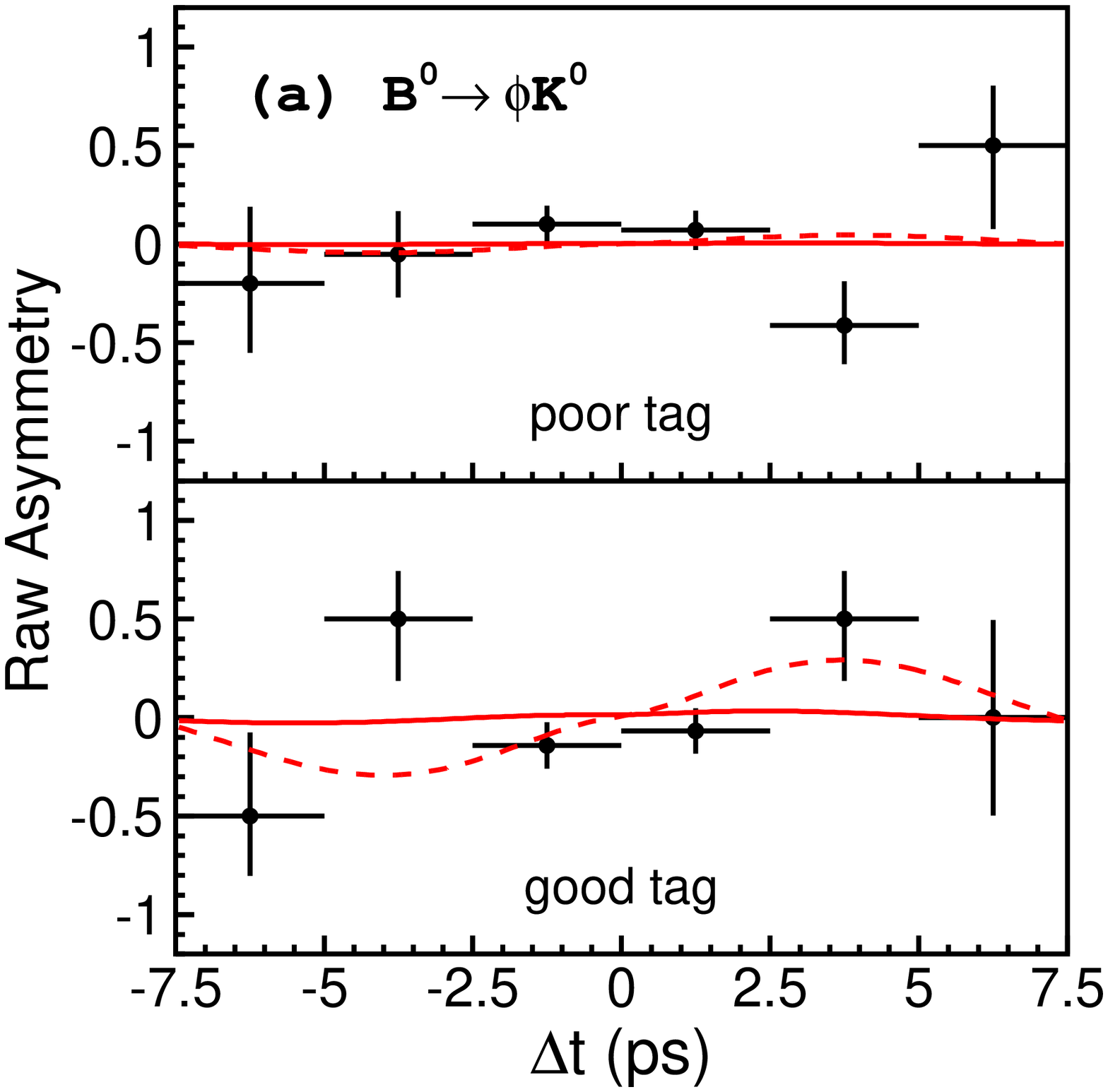}
\hspace*{-3mm}
\includegraphics[width=0.33\textwidth]{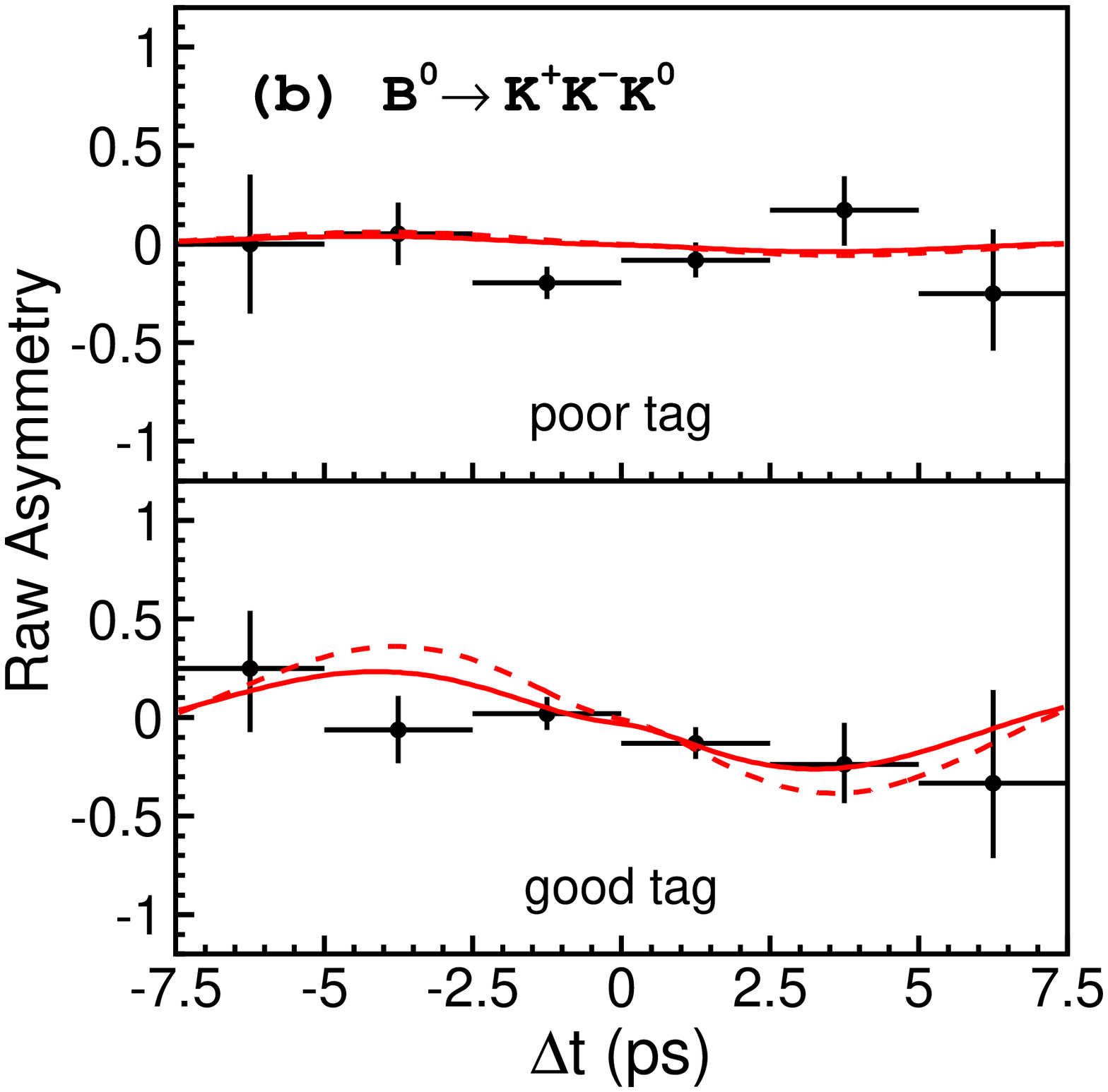}
\hspace*{-3mm}
\includegraphics[width=0.33\textwidth]{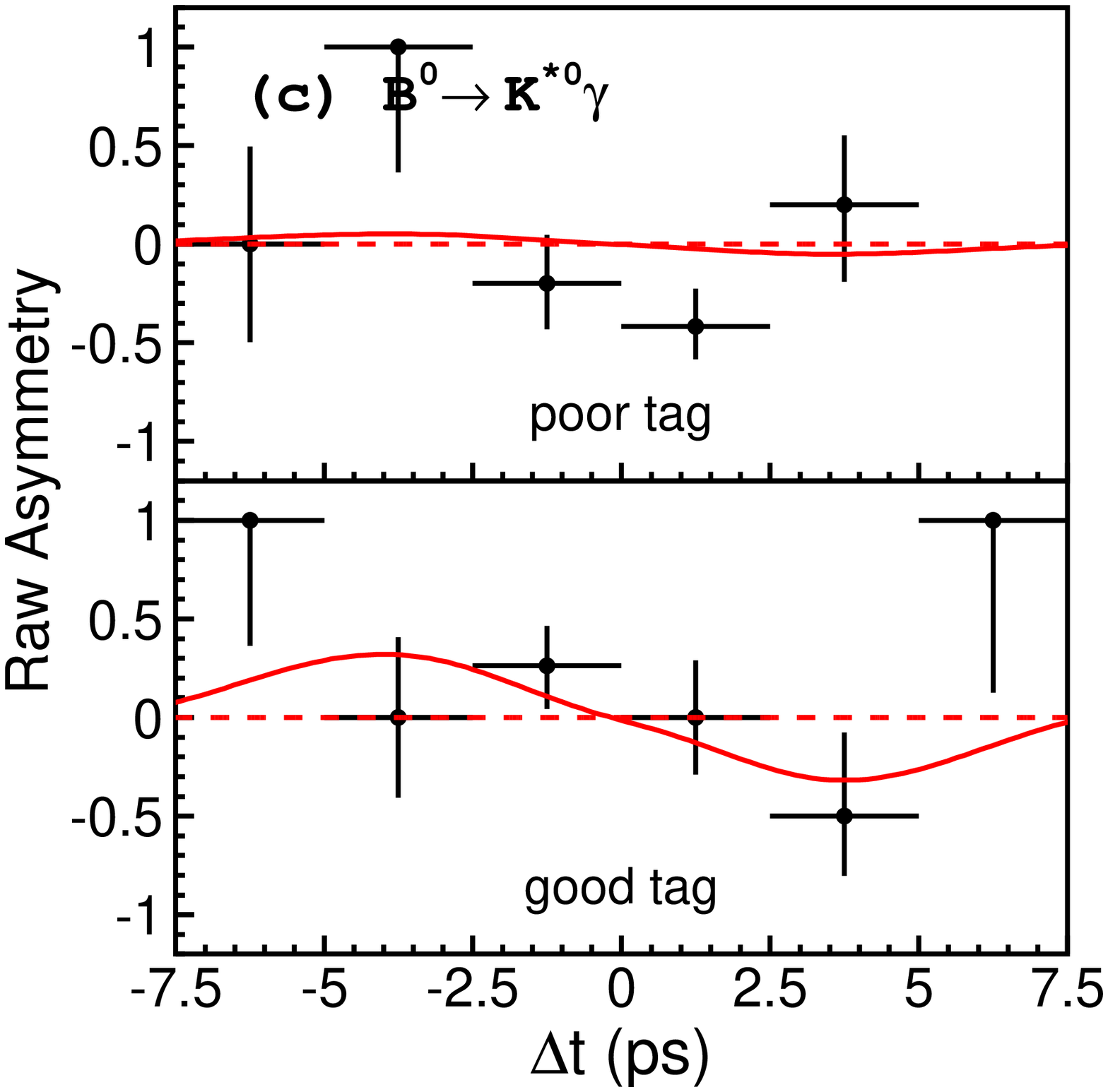}
\vspace*{0pt}
\caption{
The asymmetry, $A$, in each $\Dt$ bin for 
(a) $\bz\to\phi\kz$, 
(b) $\bz\to\kp\km\ks$ and
(d) $\bz\to\kstarz\gamma$.
The solid curves show the result of the 
unbinned maximum-likelihood fit.
The dashed curves show the SM expectation with $\sin2\phi_1$ = +0.73 
($\cals = 0$ for $\bz\to\kstarz\gamma$) and $\cala$ = 0.}
\label{fig:asym}
\end{figure}

Table \ref{tab:results} summarizes the fit results of $\cals$ and
$\cala$. We define the raw asymmetry in each $\Dt$ bin by
$(N_{q=+1}-N_{q=-1})/(N_{q=+1}+N_{q=-1})$, where $N_{q=+1(-1)}$ is the
number of observed candidates with $q=+1(-1)$~\cite{footnote:phikz}.
Figures~\ref{fig:asym}(a-c) show the raw asymmetries for some channels.
Note that these projections onto the $\Delta t$ axis do not take into
account event-by-event information (such as the signal fraction, the
wrong tag fraction and the vertex resolution), which is used in the
unbinned maximum-likelihood fit.

Various crosschecks of the measurement are performed.
We reconstruct charged $B$ meson decays
that are the counterparts of the $\bz\to\fCP$ decays
and apply the same fit procedure.
All results for the $\cals$ term are consistent with no 
$CP$ asymmetry, as expected. 
Lifetime measurements are also performed for 
the $\fCP$ modes and the corresponding charged $B$ decay modes.
The fits yield
$\taubz$ and $\taubp$ values consistent with the world-average values.
MC pseudo-experiments are generated for each decay mode to
perform ensemble tests.
We find that the statistical errors obtained
in our measurements are all consistent
with the expectations from the ensemble tests.

For the $\bz\to\phi\kz$ decay,
a fit to DS-I alone yields
$\cals = -0.68\pm 0.46$(stat) and 
$\cala = -0.02 \pm 0.28$(stat),
while a fit to DS-II alone yields
$\cals = +0.78\pm 0.45$(stat) and $\cala = +0.17 \pm 0.33$(stat).
Note that the results for DS-I
differ from our previously published results~\cite{Abe:2003yt}
$\cals = \SphiksResPrv$ and $\cala = \AphiksResPrv$,
as decays $\bz\to\phi\kl$ and $\phi\ks~(\ks\to\piz\piz)$ are
included in this analysis.
From MC pseudo-experiments, 
the probability that the difference between $\cals$ values in
DS-I and DS-II is larger than the observed one (1.46)
is estimated to be 4.5\%.
As all the other checks also yield
results consistent with expectations,
we conclude that the difference in $\cals_{\phi\kz}$
between the two datasets is due to a statistical fluctuation.

Averaging the result of $\cals$ measurement for all the $b\to s$
channels (except $\bz\to\kstarz\gamma$), we obtain $\sinbb = \SbsqqResult$
as a weighted average, where the error includes both statistical and
systematic errors. The result differs from the SM expectation by
2.4 standard deviations.

\begin{thebibliography}{0}
\bibitem{footnote:phikz}
We use an alternative definition of the raw asymmetry,
$\frac{N_{q\xi_f = -1}-N_{q\xi_f = +1}}{N_{q\xi_f = -1}+N_{q\xi_f = +1}}$, 
for the decay $\bz\to\phi\kz$ to take the opposite $CP$ parities
for the decays $\bz\to\phi\ks$ and $\phi\kl$ into account.
\end{thebibliography}
\end{document}